\documentclass[12pt]{iopart}
\usepackage{url}
\usepackage{graphicx}  
\begin{document}

\title[Extrasolar Planets in the Classroom]{Extrasolar Planets in the Classroom}
\author{Samuel J. George$^{1,2}$}

\address{$^1$ School of Physics \& Astronomy, University of Birmingham, Edgbaston, Birmingham B15 2TT, United Kingdom \\ 
$^2$ Institute for Space Imaging Science, University of Calgary, 2500 University Drive, Calgary, Canada, T2N 1N4}
\ead{samuel@ras.ucalgary.ca}

\begin{abstract}
The field of extrasolar planets is still, in comparison with other astrophysical
topics, in its infancy. There have been about 300 or so extrasolar planets 
detected and their detection has been accomplished by various different 
techniques. Here we present a simple laboratory experiment to show how planets are detected using the transit technique. Following the simple analysis procedure describe we are able to determine the planetary radius to be $1.27 \pm 0.20 R_{J}$ which, within errors agrees with the establish value of $1.32 \pm 0.25 R_{J}$.
\end{abstract}

Accepted for publication in: {\it Physics Education}\par

\section{Introduction}
It was only in 1991 that the first extrasolar planets were discovered. They were discovered around a pulsar, PSR1257+12, by measuring the variation in arrival times of radio pulses \cite{ref2}. The first detection of a planet around a main sequence star was in 1995. This was found around the Sun-like star 51 Peg and has a mass 0.47 that of Jupiter \cite{ref3}. Unexpectedly, it has an orbital period of 4.3 days. This is much smaller than the 4333.3 days that Jupiter takes to orbit the Sun. 

The main method used for detection of extrasolar planets is the radial velocity technique. This is also know as the Doppler wobble method \cite{ref4,ref6,ref9}. The basic idea is that the star and the planet both orbit around their common centre of gravity, obviously this is towards the star.  As the star moves towards the observer the wavelength of the light shortens 
(i.e. is blue-shifted) and when it moves away the light is red-shifted. By measuring this slight change the presence of the planet can be inferred (so not directly detected). A description of a classroom experiment using this method can be found in \cite{ref7} and a planet finding tool is described in \cite{ref8}. 

In this experiment we will concentrate on another technique - planetary transits \label{transits} \cite{ref1,ref10}. A transit of an extrasolar planet occurs when the planet passes in front of its host star in the observer's line of sight. During a transit we observe a reduction in the intensity of light from the star. This fluctuation will be small and typically of the order of one percent of the star's intensity. The principle is the same as in a solar eclipse; when the Moon passes in front of the Sun the intensity is reduced by a vast amount. This is an indirect method of detection as the shape of the light curve implies the presence of the planet. For longer period planets they must be in an orbit with an inclination very close to $90^{\circ}$, thus limiting the chances of observing a transit. Using transits to detect extrasolar planets was first suggested (along with the radial velocity method) by Otto Struve in 1952 \cite{ref5}. It was not until 1999 that the first transiting planet was observed - around HD~209458 \cite{ref1}. Observing extrasolar planetary transits has become increasingly popular since the equipment required can be very basic. There is a further complication that should at least be considered; this is limb darkening. Limb darkening refers to the reduction of the brightness of the star as one moves away from the centre towards the edge (or limb). This effect occurs due to the density and temperature of the star falling off as distance increases from the star's centre. For well defined transits this effect is negligible. The shape of the dip in the light curve is altered causing the determined planetary radius to be incorrect.

In addition, by precisely measuring the periodicity of a transiting planet one could detect an otherwise invisible planet. The gravitational perturbation will cause the transit times to deviate from simple periodicity. The variation of the times of transit is proportional to the ratio of the planetary masses. For an Earth-mass planet effecting a transiting Jupiter-mass planet the signal can be as large as a few minutes over a period of a few months to years.  

\section{The Laboratory Experiment}

We now describe a simple laboratory experiment to detect the planet orbiting around the star, HD~209458. The procedure is based upon that of laboratory aimed at 1st year undergraduates at the University of Birmingham. This version of the experiment is aimed at advanced secondary school students and we have run workshops based on these ideas with over 300 14-18 year olds, with all students producing reasonable results. The full undergraduate experiment is run over 5 hours whereas the shorter workshop can be completed in 1.5 hours.

\subsection{Aims and Objectives}

The aim of this experiment is to use observations made of the brightness of the star HD~209458 over a 7 hour period to infer the properties of the planet orbiting it. This experiment introduces to the student Kepler's laws, orbital mechanics, data handling, error analysis, and current astrophysics topics. The analysis is not complex and can all be carried out using standard software packages, such as Open Office Calc or Microsoft Excel.

\subsection{Education benefits}

This is a good practical example of astronomy in the classroom. We believe that astronomy is important to be introduced in the classroom, however, due to the nature of the subject this is difficult in an experimental manner. We suggest that this would broaden the physical knowledge of students by introducing a modern research field into the classroom. This experiment also allows for exploration of key physical concepts such as: orbits, celestial mechanics, stellar types, evolution of the solar system, life in the Universe. Outside of physics this experiment also demonstrates topics of statistics (in particular distributions - mean, minimum, standard deviation), graphing, error analysis, computing skills and data handling.

\subsection{The Data Set}
Here, we make use of ground-based data collected by the STARE (STellar Astrophysics and Research on Exoplanets) project of HD~209458. This is the data set used to first detect the extrasolar planet in this system \cite{ref1}. Data for various other transiting systems can be found on the Extrasolar Planets Encyclopedia (\url{http://exoplanet.eu}).

The observational data lists the time (in fractional days) around the centre of the transit, the normalized intensity of the star, and the $1\sigma$ error on the normalized intensity. The file can be found on-line at \url{http://www.sr.bham.ac.uk/~samuel/schools/exoplanet/} in various formats (including plain text, Microsoft Excel and Open Office formats). 

\subsection{The Size of the Planet}

\begin{figure}
\begin{center}
\includegraphics[scale=0.6]{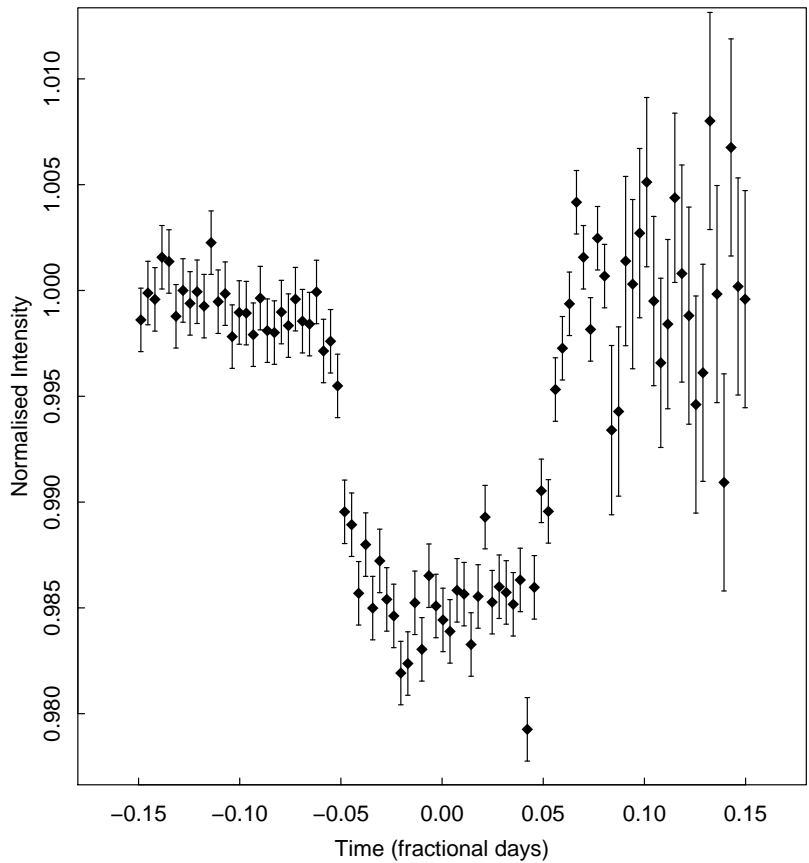}
\caption{The light curve of HD209458, one can clearly see the dip in intensity due to the planetary transit. It is also worth noting the large scatter in the data after the transit and is a useful discussion point on the reliability of the measurements. \label{dataplot}}
\end{center}
\end{figure}

To determine the size of the planet a plot of observed intensity against time should be made (see figure \ref{dataplot}). One now needs to measure the fractional depth of the transit. The fractional depth can be measured in various ways and we now describe a simple procedure. Determine when the transit starts and ends. Separate the data in time and calculate both mean ($B_{min}$) during and out of transit ($B_{max}$). The fractional depth ($\Delta F$) is determined by, 

\begin{eqnarray}
\Delta F = \frac{B_{max} - B_{min}}{B_{max}} .
\end{eqnarray}

Error bars should be added on the graph. This is an ideal time to discuss data reliability. In this case the data after the transit is rather noisy. For this experiment we assume that the planet orbiting HD~209458 crossed along the equator of the star (as seen from our perspective) but this is not necessarily true. The larger the inclination of the orbit of the planet, the shorter the transit time will be. If a planet crosses the equator of the star then this is the situation where the inclination is 90$^{\circ}$. Increasing or decreasing the inclination causes the chord across the stellar surface to be smaller and eventually we get to a point when the transit is not detected.

By measuring the depth of the dip during the transit we are able to determine the size of the planet. The transit depth is proportional to the ratio of the planet to star disc area, i.e., $(R_p / R_{*})^{2}$. If the size of the star is know (for brighter stars by direct measurement or more generally from spectral type) then the size of the planet can be determined. For the Jupiter-Sun system (radius of Jupiter, $R_{J} = 71000$~km; radius of the Sun, $R_{\odot} = 696 000$~km) the fractional depth of the dip would about 1 per cent. In the case of this experiment we take the radius of HD~209458 to be $1.1 R_{\odot}$ thus allowing us to determine the radius of the planet.

\subsection{The Size of the Orbit}

\begin{figure}[htb!]
\begin{center}
\includegraphics[scale=0.8]{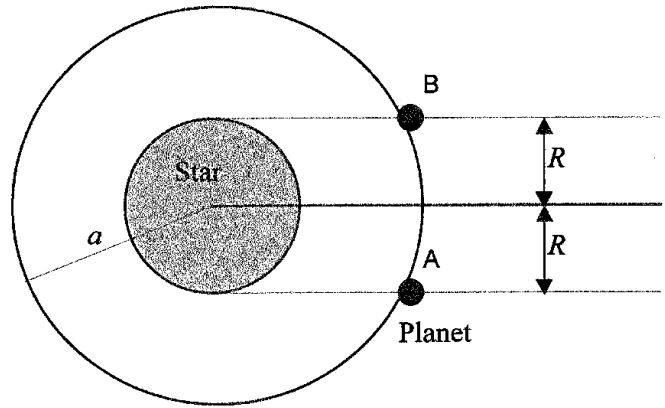}
\caption[The transit of a planet]{The transit of a planet (passing from A to B) around a star. The observer is located off the right-hand side of the paper. In this model we assume a circular orbit for the planet. \label{The transit of a planet}}
\end{center}
\end{figure}

By measuring the length of the transit we are able to determine the planetary orbital radius. Figure \ref{The transit of a planet}  shows a diagram of the transit from above. The observer is located off the right-hand side of the paper and therefore observers a transit as the planet passes from A to B. To simplify the mathematics we make the assumption that the orbit of the planet is circular. In most cases this is not true and means we underestimate the orbital period. For example Jupiter is in an elliptical orbit with eccentricity of 0.05. We also assume that orbital radius ($a$) is much larger than the stellar radius ($R_{*}$). This is true for the planets in our Solar System, but not always for extrasolar planets and results in an overestimate of the orbital period. The orbital period ($P$) is determined by,

\begin{eqnarray}
\label{Period_planet}
P = \frac{2 \pi a}{v_{c}} ,
\end{eqnarray}

where $v_{c}$ is the circular velocity of the planet. By using figure \ref{The transit of a planet} we are able to determine the transit time, $\tau$ (i.e. the time to pass from A to B) to be (using speed = distance / time), 

\begin{eqnarray}
\label{time_transit}
\tau  = \frac{2 R}{v_{c}} .
\end{eqnarray}

By combining equations \ref{Period_planet} and  \ref{time_transit} one is left with

\begin{eqnarray}
\tau  = \frac{R P}{\pi a} .
\end{eqnarray}

Using the STARE project data the transit is found to repeat some $3.5250 \pm 0.003$ days later (i.e., the time between successive transits) \cite{ref1}. Combining all the information together one can determine the orbital radius of the planet. This is much smaller than what we would expect if we assume the Solar System to be typical.  Jupiter has an orbital radius of 5.2 AU whilst HD~209458b is 0.045 AU. HD~209458b falls into the class of extrasolar planets known as ``hot Jupiters''. These are planets of similar size to Jupiter but orbiting closer to the host star.

\subsection{The Mass of HD~209458b}

By applying gravitational physics one can derive an estimate of the mass of the star ($M_{*}$). This is achieved by using Kepler's 3rd law (the square of the orbital period of a planet is directly proportional to the cube of the semi-major axis of its orbit), 

\begin{eqnarray}
P  = \frac{2 \pi}{\sqrt{G M_{*}}} a^{3/2}, 
\end{eqnarray}

where, G is the universal gravitational constant ($6.67 \times 10^{-11}$m$^{3}$kg$^{-1}$s$^{-2}$).

\subsection{Results}
By following this analysis we find the following:

\begin{enumerate}
\item{Fractional Depth $= 0.0141 \pm 0.0004$}
\item{Radius of planet $= 1.27 \pm 0.20 R_{J}$}
\item{Orbital Radius $= 0.047 \pm 0.001$~AU}
\item{Mass of the star $= 1.06 \pm 0.01 M_{\odot}$}
\end{enumerate}

These results compare well with the currently accepted values; radius of planet $= 1.32 \pm 0.25 R_{J}$, orbital radius $= 0.045 \pm 0.001$~AU, mass of the star $1.01 (\pm 0.066) M_{\odot}$.

\section{Experiences}

This procedure has been carried out with many different groups of students and it seems that a 1 hour session was enough time to give brief insights into the experiment. A 2 hour session would allow for all students to fully complete the experiment. When this experiment is run with undergraduate students (with significantly less information and more vigorous data analysis techniques required) they have 5 hours. In a 2 hour session the more able students will have time for further discussion and a more vigorous approach to the removal of ``bad data''. Some limitations are clear. Students who are not familiar with spreadsheet software (such as Microsoft Excel or Open Office Calc) do struggle with the initial data manipulation (though the experiment can be done with printed out graphs and a ruler). The concept of error bars confuses students and it is best if experimental error is recapped in advance of the experiment starting. Students also have a problem visualising the experiment so a model, such as an orrery and a torch, would be very helpful. 

\section{Summary}

In conclusion we have presented an astronomy laboratory experiment that can be undertaken in the classroom with little equipment required. The results obtained are consistent with more robust analysis methods. This experiment provides a simple way of introducing orbital dynamics and extrasolar planets to students. It also introduces basic data analysis techniques. This experiment has been followed, successfully, by a range of students from undergraduate to pre-GCSE.

\ack
This experiment is based on an idea that was introduced into the undergraduate physics laboratories at the University of Birmingham by William Chaplin and we thank him for useful discussions on the manuscript. We also wish to thank the STARE project for placing their data online.
\\\\
This is an author-created, un-copyedited version of an article accepted for publication in Physics Education. IOP Publishing Ltd is not responsible for any errors or omissions in this version of the manuscript or any version derived from it. The definitive publisher authenticated version will be available online at \url{http://iopscience.iop.org/0031-9120}.

\end{document}